\documentclass[12pt]{article}
\usepackage{a4wide,epsfig,psfrag,amsmath,amssymb,cite,scalefnt}
\usepackage{color}

\newcommand{\apiSU}{\frac{\alpha^{\rm SU(5)}(\mu)}{4 \pi}}

\newcommand{\mgut}{M_{\rm G}}
\newcommand{\mhc}{M_{\rm H_c}}
\newcommand{\mhcp}{M_{\rm H_c^\prime}}

\newcommand{\mugut}{\mu_{\rm GUT}}
\newcommand{\mususy}{\mu_{\rm SUSY}}

\newcommand{\abbrev}{\scalefont{.9}}
\newcommand{\drbar}{$\overline{\mbox{\abbrev DR}}$}
\newcommand{\msbar}{$\overline{\mbox{\abbrev MS}}$}

\newcommand{\dred}{{\abbrev DRED}}

\newcommand{\mbMSbar}{m_{b}^{\overline{\rm MS}}}

\begin{document}

%- {{{ title, abstract:

\title{\vskip-3cm{\baselineskip14pt
    \begin{flushleft}
      \normalsize SFB/CPP-10-76\\
      \normalsize TTP10-34
  \end{flushleft}}
  \vskip1.5cm
  Minimal Supersymmetric SU(5) and Gauge Coupling Unification at Three Loops
}

\author{
  W.~Martens, 
  L.~Mihaila, 
  J.~Salomon, 
  M.~Steinhauser
  \\[1em]
  {\small\it Institut f{\"u}r Theoretische Teilchenphysik}\\
  {\small\it Karlsruhe Institute of Technology (KIT)}\\
  {\small\it 76128 Karlsruhe, Germany}
}

\date{}

\maketitle

\thispagestyle{empty}

\begin{abstract}
  We consider the relations between the gauge couplings at the
  electroweak scale and the high scale where unification of the three
  gauge couplings is expected.
  Threshold corrections are incorporated both at the supersymmetric
  and at the grand unified scale and, where available three-loop
  running and two-loop decoupling are employed.
  We study the impact of the current experimental uncertainties of the
  coupling constants and the supersymmetric mass spectrum on
  the prediction of the super-heavy masses within the so-called minimal
  supersymmetric SU(5). 
  As a main result of the three-loop analysis we confirm that
  minimal supersymmetric SU(5) cannot be ruled out by the current
  experimental data on proton decay rates.

  \medskip


\end{abstract}

%- }}}

\newpage

%- {{{ Introduction:

\section{\label{sec::intro}Introduction}

An appealing hint in favour of supersymmetry (SUSY) is the apparent 
unification of 
gauge couplings at a scale of about
$10^{16}$~GeV~\cite{Ellis:1990wk,Amaldi:1991cn,Langacker:1991an}. In
particular, the gauge coupling unification is predicted,
even under the assumption of a
minimal particle content of the underlying Grand Unified Theory (GUT), the
so-called minimal SUSY SU(5) model~\cite{Dimopoulos:1981zb,Sakai:1981gr}.
An important feature of
this model is that the requirement of minimality renders it  the most
predictive model from the currently known candidates for 
GUTs. However, immediately after the formulation of the minimal SUSY SU(5) it
has been noticed that within SUSY GUTs new dimension-five operators
cause a rapid proton decay~\cite{Sakai:1981pk,Weinberg:1981wj}.
This aspect was intensively studied over 
the last thirty years with the extreme conclusions
of Refs.~\cite{Goto:1998qg,Murayama:2001ur} that the minimal SUSY SU(5)
model is ruled
out by the combined constraints from proton decay and gauge coupling
unification.
However, in the following years several careful analyses have shown that
the constraint on the coloured
Higgs triplet mass from proton decay was actually too strong, and that
minimal SUSY SU(5) is still a perfectly viable theory.
The proton decay rate for the dominant channel $p\rightarrow
K^+\overline{\nu}$ can be suppressed either
by sfermion mixing~\cite{Bajc:2002bv} or by taking into account higher
dimensional operators induced at the Planck
scale~\cite{EmmanuelCosta:2003pu,Wiesenfeldt:2004qz,Bajc:2002pg}. 
Such operators can also lead to a successful down quark and
charged lepton Yukawa coupling
unification for the first two families, which is not possible in the
renormalizable version of the minimal
SUSY SU(5). Planck-scale-induced operators also change 
the boundary conditions for the gauge couplings at the GUT scale, since in
their presence the colour octet and the isospin
triplet Higgs bosons contained in the
$\mathbf{24}_H$ representation do not necessarily 
have the same mass anymore, as it is the case in
the minimal model. This fact allows for significantly weaker constraints 
from gauge coupling unification
on the colour triplet Higgs boson
than in the renormalizable model.

Recently, new experimental data for the relevant input
parameters~\cite{Kobayashi:2005pe,Amsler:2008zzb,Bethke:2009jm}
and substantial progress on the theory
side~\cite{Ferreira:1996ug,Harlander:2005wm,Bauer:2008bj,Harlander:2009mn}
became available.
This encourages us to reanalyze minimal SUSY SU(5) focusing on the
aspect of gauge coupling unification. 
 We emphasize that in our analysis we restrict ourselves to the renormalizable
 version of minimal SUSY SU(5) although in this way the unification of
 the first and second generation Yukawa couplings can not be
 achieved. In our analysis we ignore this
 fact since it is not relevant here.
 The constraints on the mass of the coloured Higgs triplet that are
 derived can easily be translated to the non-renormalizable version of
 minimal SUSY SU(5) by comparison of, e.g., Eqs.~(9) and~(10)
of Ref. \cite{Bajc:2002pg} and Eq.~(\ref{eq::gutrel}) in our paper.
In this sense, our results represent the ``worst case scenario''.

More precisely, we review in this paper the constraints on the mass of the
coloured Higgs 
triplet $\mhc$ and the grand unification scale\footnote{See 
  Section~\ref{sec::rge} for the exact definition of $\mhc$ and
  $\mgut$.} $\mgut$ 
within minimal SUSY
SU(5)~\cite{Dimopoulos:1981zb,Sakai:1981gr} taking into
account the latest experimental data for the weak scale parameters and
the most precise theoretical predictions currently available. 
We adopt the renormalization group method in the ``bottom-up'' approach
to predict the values of the two parameters and take into
account threshold corrections generated by the superpartners of the
SM particles as well as those due to  the super-heavy SUSY-GUT particles.
In this way, we can derive the two SUSY-GUT parameters from the
knowledge of the gauge coupling constants of the SM at the electroweak
scale and the minimal supersymmetric SM (MSSM) mass spectrum.
In addition, we implement the perturbativity restrictions
for all gauge, Yukawa and Higgs self couplings, {\it i.e.}, we require that they 
are smaller than one up to the Planck scale. 

Besides the minimal SU(5) we also briefly discuss the phenomenological
consequences of the Missing Doublet Model (MDM) which has been designed in
order to avoid unnatural
doublet-triplet splitting in the Higgs fields of the $\mathbf{5}$ 
and $\mathbf{\bar{5}}$ representations.

The remainder of the paper is organized as follows: In the next
Section we introduce our framework and describe the tools we have used
for our analysis. In particular we specify the underlying GUT theory
and describe in detail our procedure for the running and decoupling.
In Section~\ref{sec::pheno} the phenomenological consequences are
discussed where we study in particular the constraints on the GUT
masses of minimal SUSY SU(5). Finally, we present our conclusions in
Section~\ref{sec::concl}.

%- }}}
%- {{{ Theoretical framework:

\section{\label{sec::rge}Theoretical framework}

%- {{{ Minimal SU(5):

\subsection{Minimal SU(5) and Missing Doublet Model}

In Section~\ref{sec::pheno} we discuss in
detail the restrictions on ``minimal SU(5)'', however, mention also
briefly the consequences for the so-called ``Missing Doublet Model''.
For convenience of the reader we introduce in this Subsection some
details on these models.

The superpotential of minimal SU(5) \cite{Dimopoulos:1981zb} is given by
\begin{eqnarray}
  {\cal W} &=& M_{1}\rm{Tr}(\Sigma^2)+\lambda_1 \rm{Tr}(\Sigma^3)+
  \lambda_2 \bar{H}\Sigma H +M_2 \bar{H} H \nonumber\\
  &&\mbox{}+ \sqrt{2} Y_d^{ij}\Psi_i \phi_j \bar{H} + \frac{1}{4} Y_u^{ij} \Psi_i\Psi_j H
  \,,
\end{eqnarray}
where $\Psi_i$ and $\phi_i$ ($i=1,2,3$ is a generation index) are matter
multiplets in the $\mathbf{10}$- and $\mathbf{\overline{5}}$-dimensional 
representation of SU(5), respectively, and the field $H$ ($\bar{H}$) is
realized in the $\mathbf{5}$ ($\mathbf{\bar{5}}$) representation.
SU(5) is broken to SU(3)$\times$SU(2)$\times$U(1) if 
the adjoint Higgs boson $\Sigma\equiv\Phi^a T^a$ ($a=1,\ldots,24$) gets the
vacuum expectation value  $\langle\Sigma\rangle =
V/(2\sqrt{30})\times\rm{diag}(-2,-2,-2,3,3)$, 
with $V=-4\sqrt{30} M_1/(3\lambda_1)$. 
Choosing $\langle\bar{H}\rangle=\langle H\rangle\ll V$ and in addition
 imposing the (tree-level-)fine-tuning
condition $M_2=-\sqrt{3}\lambda_2 V/\sqrt{40}$  the isodoublets in $H$ and
$\bar{H}$ remain massless.
Furthermore, one gets the following super-heavy mass spectrum:
\begin{eqnarray}
  M_X^2=\frac{5}{12}g^2V^2\, ,\quad M_{H_c}^2 =
  \frac{5}{24}\lambda_2^2 V^2\,,\quad M_{\Sigma}^2\equiv 
  M^2_{(8,1)} = M^2_{(1,3)} = 25 M^2_{(1,1)} = \frac{15}{32}\lambda_1^2 V^2 
  \,,
\end{eqnarray} 
where the indices in round brackets refer to the SU(3) and SU(2) 
quantum numbers.
Here $M_{\Sigma}$ denotes the mass of the colour octet part of the
adjoint Higgs boson $\Sigma$ and $M_{H_c}$ stands for the mass of the colour
triplets of $H$ and $\bar{H}$, $M_X$ is the mass of the gauge bosons
and $g$ is the gauge coupling.
The equality $ M^2_{(8,1)} = M^2_{(1,3)}$ only holds 
if one neglects operators that are suppressed by $1/M_{\rm Pl}$ as we do here.
We emphasize again that taking into account such operators can
considerably weaken the constraint
on the colour triplet mass $M_{H_c}$ that is derived later in this
 paper~\cite{EmmanuelCosta:2003pu,Wiesenfeldt:2004qz,Bajc:2002pg}.
However, the altered constraint can easily be derived from our results.

The effects of the super-heavy particle masses on the MSSM gauge
couplings can be parametrized with the help of decoupling 
coefficients. If we interpret the MSSM as the low-energy effective
theory of the SUSY SU(5) model, we can define its three gauge coupling
constants  as functions of the unique  SU(5) gauge
coupling $\alpha^{\rm SU(5)}$ through
\begin{eqnarray}
  \alpha_i^{\rm MSSM}(\mu_{\rm GUT}) = \zeta_{\alpha_i}(\mu_{\rm
    GUT},\alpha^{\rm SU(5)}, M_{H_c}, M_{X},
  M_{\Sigma})\, \alpha^{\rm SU(5)}(\mu_{\rm GUT})\,, \quad i=1,2,3\,,
  \label{eq::dec}
\end{eqnarray}
where $\zeta_{\alpha_i}(\mu_{\rm GUT},\alpha^{\rm SU(5)}, M_{H_c},
M_{X},M_{\Sigma})$ denote 
the decoupling coefficients that can be obtained from Green's functions
with external light particles computed in the full and effective theory.
The scale $\mu_{\rm GUT}$ is an unphysical parameter, not fixed by theory.
The dependence of physical observables on $\mu_{\rm GUT}$ thus provides an
estimation of the theoretical uncertainties within fixed order
perturbation theory.

For SUSY theories the most convenient
regularization scheme is Dimensional Reduction (\dred{})\cite{Siegel:1979wq}
which we also adopt in our
calculation. As a consequence, the coupling constants appearing in
Eq.~(\ref{eq::dec}) are renormalized minimally  in the so-called
\drbar{} renormalization scheme.
The one-loop formulas of the decoupling coefficients for a 
general gauge group have been known for a long
time~\cite{Hall:1980kf,Weinberg:1980wa,Einhorn:1981sx,Marciano:1981un}. 
The specification to
minimal SUSY SU(5) reads~\cite{Hagiwara:1992ys,Dedes:1996wc}
\begin{eqnarray}
  \zeta_{\alpha_1}(\mu)&=&1+\apiSU\left(-\frac{2}{5}\,L_{\mu H_c}+10\,L_{\mu
      X}  
  \right)
  \,,
  \nonumber\\
  \zeta_{\alpha_2}(\mu)&=& 1+\apiSU\left(-2\,L_{\mu\Sigma}+  6\,L_{\mu X} \right)
  \,,
  \nonumber\\
  \zeta_{\alpha_3}(\mu)&=& 1+\apiSU\left(-L_{\mu
      H_c}-3\,L_{\mu\Sigma}+4\,L_{\mu X} 
  \right)\,,
  \label{eq::gutdec}
\end{eqnarray}
where $L_{\mu x}=\ln(\mu^2/M_x^2)$ and for simplicity  we  keep
from the list of arguments of the coefficients $\zeta_{\alpha_i}$ only
the decoupling scale. 

The Missing Doublet Model~\cite{Masiero:1982fe,Grinstein:1982um} is designed to avoid unnatural 
doublet-triplet splitting in the field $H$ that is present in the minimal model.
This is achieved at the cost of introducing additional Higgs
fields $\Theta$ and $\bar{\Theta}$ in the large SU(5) 
representations $\mathbf{50}$ and $\overline{\mathbf{50}}$ that do not contain 
any isodoublets and thus only couple to the colour triplets in $H$ and $\bar{H}$.
To break SU(5) another Higgs field $\Sigma$ in the
 $\mathbf{75}$ representation is used instead of the $\mathbf{24}$ as 
in the minimal model. The superpotential reads
\begin{eqnarray}
  {\cal W} &=& M_{1}\rm{Tr}(\Sigma^2)+\lambda_1 \rm{Tr}(\Sigma^3)+
  \lambda_2 H\Sigma \Theta + \bar{\lambda}_2
 \bar{H}\Sigma \bar{\Theta} +M_2 \bar{\Theta} \Theta \nonumber\\
  &&\mbox{}+ \sqrt{2} Y_d^{ij}\Psi_i \phi_j \bar{H} + \frac{1}{4} Y_u^{ij} \Psi_i\Psi_j H
  \,.
\end{eqnarray}
After $\Sigma$ develops a vacuum expectation value the spectrum of the theory
can be parametrized by five mass
 parameters~\cite{Dedes:1996wc} $M_X, M_{H_c}, M_{H_{c'}}, M_\Sigma$ and $M_2$.
The last is assumed to be of $\mathcal{O}(M_{\rm Pl})$ so that the representations  
$\mathbf{50}$ and $\overline{\mathbf{50}}$ do not contribute to the running
above the unification scale and below $M_{\rm Pl}$. 
Otherwise, due to large group factors of these representations,
the perturbativity requirement cannot be fulfilled.
In this case the decoupling constants read~\cite{Hagiwara:1992ys}\footnote{The occurrence
  of the last term
  in the round brackets of each equation  is due to the use of relations
  between the super-heavy masses.}
\begin{eqnarray}
  \zeta_{\alpha_1}(\mu)&=&1+\apiSU\left(-\frac{2}{5}\,L_{\mu H_c}
    -\frac{2}{5}\,L_{\mu H_{c'}}    
    +10\,L_{\mu X}-20\,L_{\mu \Sigma} + 10 \ln\frac{64}{625}  \right)
  \,,\nonumber\\
  \zeta_{\alpha_2}(\mu)&=& 1+\apiSU\left(-22\,L_{\mu\Sigma}+  6\,L_{\mu X} +
    6\ln\frac{4}{25} \right) 
  \,,\nonumber\\
  \zeta_{\alpha_3}(\mu)&=& 1+\apiSU\left(-L_{\mu H_c}-L_{\mu H_{c'}}
    -23\,L_{\mu\Sigma}+4\,L_{\mu X} + 4\ln\frac{64}{78125}
  \right)\,.
  \label{eq::gutdec2}
\end{eqnarray}
As we will see later, the presence of the large representation
$\mathbf{75}$ in this model leads to huge theoretical
uncertainties due to the variation of the unphysical scale $\mu_{\rm GUT}$.

%- }}}
%- {{{ Running and decoupling:

\subsection{\label{sec::rundec}Running and decoupling}

It is well known that gauge coupling
 unification is highly sensitive to the super-heavy mass
spectrum~\cite{Hisano:1992mh}. This property
allows us to probe unification through precision measurements of
low-energy parameters like the gauge couplings at the electroweak
scale or the supersymmetric mass spectrum.
A simple algebraic exercise taking into account
the naive step-function approximation~\cite{Ross:1992tz}
based on one-loop RGEs provides analytical formulas for the determination
of the 
GUT spectrum as a function of the three gauge couplings measured at the
$Z$-boson mass scale. An estimate of three-loop as compared to two-loop
running has been obtained in Ref.~\cite{Jack:2004ch}.
Note, however, that a consistent treatment requires the
implementation of  $n$-loop RGEs and $(n-1)$-loop threshold
corrections. We have adopted this approach for $n=1, 2$ and $3$,
whenever the required theoretical input was available, 
and have solved numerically the system of differential equations.

Crucial input for our analysis constitute the precise values of the
gauge couplings at the electroweak scale. They are obtained from the 
weak mixing angle in the $\overline{\rm MS}$
scheme~\cite{Amsler:2008zzb}, the QED coupling constant at zero
momentum transfer and its hadronic~\cite{Teubner:2010ah}
contribution in order to obtain its counterpart at the $Z$-boson scale,
and the strong
coupling constant~\cite{Bethke:2009jm}.\footnote{We adopt the central
  value from Ref.~\cite{Bethke:2009jm}, however, use as our default
  choice for the uncertainty $0.0020$ instead of $0.0007$.} 
The corresponding central values and uncertainties 
read
\begin{eqnarray}
  \sin^2\Theta^{\overline{\rm MS}} &=& 0.23119 \pm 0.00014
  \,,\nonumber\\
  \alpha &=& 1/137.036
  \,,\nonumber\\
  \Delta\alpha^{(5)}_{\rm had} &=& 0.02761\pm 0.00015
  \,,\nonumber\\
  \alpha_s(M_Z) &=& 0.1184 \pm 0.0020
  \,.
  \label{eq::parin5}
\end{eqnarray}

Whereas $\sin^2\Theta^{\overline{\rm MS}}$ and $\alpha_s(M_Z)$ are
already defined in the $\overline{\rm MS}$ scheme,
$\Delta\alpha^{(5)}_{\rm had}$ constitute corrections to the on-shell
value of $\alpha$. 
In order to obtain the corresponding $\overline{\rm
  MS}$ result we add the leptonic~\cite{Steinhauser:1998rq} and top
quark~\cite{Kuhn:1998ze} contribution, 
$\Delta\alpha^{(5)}_{\rm lep} = 314.97686\times 10^{-4}$ and
$\Delta\alpha^{(5)}_{\rm top} = (-0.70 \pm 0.05) \times 10^{-4}$,
and apply the transition formula to the \msbar{} scheme~\cite{Amsler:2008zzb}
\begin{eqnarray}
  \Delta\alpha^{(5),\overline{\rm MS}}
  - \Delta\alpha^{(5),\rm OS} 
  &=&
  \frac{\alpha}{\pi}\left(\frac{100}{27}-\frac{1}{6}-\frac{7}{4}\ln\frac{M_Z^2}{M_W^2}
  \right)
  \,\,\approx\,\, 0.0072
  \,.
\end{eqnarray}
This leads to
\begin{eqnarray}
  \alpha^{\overline{\rm MS}}(M_Z) &=& \frac{\alpha}{1-\Delta\alpha_{\rm lep}^{(5)} -
    \Delta\alpha^{(5)}_{\rm had} - \Delta\alpha^{(5)}_{\rm top} - 0.0072}
  \,\,=\,\, \frac{1}{127.960 \pm 0.021}  \,.
\end{eqnarray}

In the quantities $\sin^2\Theta^{\overline{\rm MS}}$,
$\alpha^{\overline{\rm MS}}(M_Z)$ and $\alpha_s(M_Z)$ the top quark is
still (partly) decoupled. Thus, in a next step we compute the
six-flavour SM quantities using the
relations~\cite{Amsler:2008zzb,Fanchiotti:1992tu} 
\begin{eqnarray}
  \alpha^{(6),\overline{\rm MS}} &=& 
  \alpha^{\overline{\rm MS}}\left\{
    1 + \frac{4}{9}\frac{\alpha^{\overline{\rm MS}}}{\pi}
    \left[\ln\frac{M_Z^2}{M_t^2}\left(1+\frac{\alpha_s}{\pi}+
      \frac{\alpha^{\overline{\rm MS}}}{3\pi}
      \right)
      + \frac{15}{4}\left(\frac{\alpha_s}{\pi}+
      \frac{\alpha^{\overline{\rm MS}}}{3\pi}
      \right)
      \right]
    \right\}
    \,,\nonumber\\
    \sin^2\Theta^{(6),\overline{\rm MS}} &=& \sin^2\Theta^{\overline{\rm MS}}
    \left\{
    1 + \frac{1}{6} \frac{\alpha^{\overline{\rm MS}}}{\pi}
    \left(\frac{1}{\sin^2\Theta^{\overline{\rm MS}}} - \frac{8}{3}\right)
    \left[\left(1+\frac{\alpha_s}{\pi}\right)\ln\frac{M_t^2}{M_Z^2}
      - \frac{15}{4}\frac{\alpha_s}{\pi} \right]
    \right\}
    \,,
    \nonumber\\
\end{eqnarray}
where all couplings are evaluated at the scale $\mu=M_Z$.
We obtain\footnote{Since we aim for gauge couplings at the electroweak
  scale with highest possible precision we use
  four-loop running and three-loop decoupling as implemented in {\tt
    RunDec}~\cite{Chetyrkin:2000yt} 
  in order to obtain $\alpha_s^{(6)}$
  from $\alpha_s(M_Z)\equiv\alpha_s^{(5)}(M_Z)$. At such high order in
  perturbation theory there is practically no dependence on the decoupling
  scale.} 
\begin{eqnarray}
  \alpha^{(6),\overline{\rm MS}}(M_Z) &=& 1/(128.129\pm 0.021)
  \,,\nonumber\\
  \sin^2\Theta^{(6),\overline{\rm MS}}(M_Z) &=& 0.23138 \pm 0.00014
  \,,\nonumber\\
  \alpha_s^{(6)}(M_Z) &=& 0.1173\pm 0.0020
  \,.
  \label{eq::alphasin}
\end{eqnarray}
These quantities are related to the three gauge couplings via the equations
\begin{eqnarray}
  \alpha_1 &=& \frac{5}{3}
  \frac{\alpha^{(6),\overline{\rm MS}}}{\cos^2\Theta^{(6),\overline{\rm MS}}}
  \,,\nonumber\\
  \alpha_2 &=& \frac{\alpha^{(6),\overline{\rm
        MS}}}{\sin^2\Theta^{(6),\overline{\rm MS}}}
  \,,\nonumber\\
  \alpha_3 &=& \alpha_s^{(6)}
  \,,
  \label{eq::alpha123}
\end{eqnarray}
which holds for any renormalization scale $\mu$. 

To the accuracy we are aiming at one has to worry about supersymmetric
effects influencing the extraction of the couplings in
Eq.~(\ref{eq::alphasin}) from experimental data.
Due to the presence of the weak gauge bosons in the loop corrections the
weak mixing angle receives the numerically largest contributions whereas the
influence of supersymmetric particles on the electromagnetic and strong
coupling can be neglected. 

The procedure to incorporate the supersymmetric
effects on the numerical value of $\sin^2\Theta^{(6),\overline{\rm MS}}(M_Z)$
is as follows: In a first step we transfer 
$\sin^2\Theta^{(6),\overline{\rm MS}}(M_Z)$ from Eq.~(\ref{eq::alphasin})
to the \drbar{} scheme~\cite{Martin:1993yx} and apply afterwards the
supersymmetric one-loop corrections evaluated in
Ref.~\cite{Dedes:1998hg} relating the weak mixing angle in the SM to the one
in MSSM. In a next step we decouple the supersymmetric
particles~\cite{Yamada:1992kv} and finally go back to the \msbar{} scheme. 
As a result we obtain $\sin^2\Theta^{(6),\overline{\rm MS}}(M_Z)$ at the scale
$\mu=M_Z$ including virtual MSSM contributions. Note that by construction
these corrections are suppressed by the square of the supersymmetric mass scale.
We anticipate that for typical supersymmetric benchmark scenarios the
influence of supersymmetric corrections to
$\sin^2\Theta^{(6),\overline{\rm MS}}(M_Z)$ can lead to shifts in $\mhc$ 
which are of the order of 10\%. For the {\tt mSUGRA} parameters in
Eq.~(\ref{eq::msugra_par1}) the shift of 
$\sin^2\Theta^{(6),\overline{\rm MS}}(M_Z)$
amounts to about $1.4\cdot 10^{-5}$.

In addition to
the input values for the gauge coupling constants, we also need 
the $W$- and $Z$-boson pole masses $M_W$ and $M_Z$,  
the top-quark and tau-lepton pole
masses $M_t$ and $M_\tau$ and the running bottom-quark mass $\mbMSbar$.
For the convenience of the reader we also specify their numerical
values~\cite{Amsler:2008zzb,:2009ec,Chetyrkin:2009fv}
\begin{eqnarray}
  M_W &=& 80.398~\mbox{GeV}  \,,\nonumber\\
  M_Z &=& 91.1876~\mbox{GeV}  \,,\nonumber\\
  M_t &=& 173.1~\mbox{GeV}  \,,\nonumber\\
  M_\tau &=& 1.77684~\mbox{GeV}  \,,\nonumber\\
  \mbMSbar(\mbMSbar) &=& 4.163~\mbox{GeV}  \,.
\end{eqnarray}
The corresponding uncertainties are not important for our analysis.

In the following we describe in detail the individual steps of the running analysis
 needed for the  energy evolution of the gauge couplings from the scale
$\mu=M_Z$ to the unification scale.
\begin{enumerate}
\item Running within the SM from $\mu=M_Z$ to the
  SUSY scale $\mususy$.\\
  Starting from Eqs.~(\ref{eq::alphasin}) and~(\ref{eq::alpha123})
  (and adding supersymmetric effects as discussed above)
  we use the three-loop beta function of QCD~\cite{Tarasov:1980au,Larin:1993tp}
  and the two-loop RGEs in the electroweak
  sector~\cite{Ford:1992pn,Machacek:1983tz,Machacek:1983fi} in order
  to obtain the values of the gauge couplings at $\mususy\approx 1$~TeV.
  We take into account the tau, bottom and top Yukawa couplings 
  and thus solve a coupled system of six differential equations.
  Since the quartic Higgs coupling $\lambda$ enters the equations of the
  Yukawa couplings
  starting from two-loop order only we neglect its contribution.
  At this point we want to stress that $\mususy$ is not fixed but kept
  as a free parameter in our setup.

\item SUSY threshold corrections.\\
  For energies of about $\mususy\approx 1$~TeV the SUSY particles
  become active and the proper matching between the SM and the MSSM has to be
  performed. We decouple all heavy non-SM particles simultaneously at the
  scale $\mususy$ using the one-loop relations for $\alpha_1$ and $\alpha_2$
  and the Yukawa couplings from Refs.~\cite{Yamada:1992kv}
  and~\cite{Pierce:1996zz}, respectively. The SUSY-QCD decoupling effects for
  $\alpha_3$ and $m_b$ are known to two-loop order and have been computed in
  Refs.~\cite{Harlander:2005wm,Bauer:2008bj}. 
  The simultaneous decoupling might be problematic in case there is a huge
  splitting among the SUSY masses. In that case a step-by-step decoupling
  would be preferable (see, e.g., Ref.~\cite{Box:2008xu}), however, a two-loop
  calculation in that framework is still missing. 
  Furthermore, the mass splitting in almost all benchmark scenarios currently
  discussed in the literature is rather small.

  As pointed out before, a fully consistent approach would require
  two-loop threshold corrections not only in the strong but also in
  the electroweak sector. They are  not yet available, however, we
  also expect that their numerical impact is relatively small.
  Furthermore, consistency with the RG running would require mixed
  QCD-Yukawa corrections for $\alpha_3$ which are not yet available. Since the
  effect of these kind of corrections on the (three-loop) running is 
  numerically small we expect that also the impact on the decoupling is 
  small.

  In our numerical analysis we generate the SUSY mass spectrum with the help
  of the program {\tt SOFTSUSY}~\cite{Allanach:2001kg} and study the various 
  SPS (Snowmass Points and Slopes) scenarios~\cite{Allanach:2002nj,AguilarSaavedra:2005pw}.

  At this stage also the change of renormalization scheme
  from \msbar{} to \drbar{} has to be taken into account. We employed the
  one-loop conversion relations~\cite{Martin:1993yx,Harlander:2007wh} for all
  parameters except $\alpha_3$ and $m_b$ where two-loop
  relations~\cite{Harlander:2007wh,Mihaila:2009bn} have been used
  in order to be consistent
  with the decoupling at the SUSY scale.

\item Running within the MSSM from $\mususy$ to the high-energy
  scale $\mugut$.\\
  We use the
  three-loop RGEs of the MSSM~\cite{Ferreira:1996ug,Harlander:2009mn} to evolve the
  gauge and Yukawa couplings  from $\mususy$ to some very high scale
  of the order of $10^{16}$~GeV, that we denote by 
  $\mugut$, where we expect that SUSY-GUT particles become
  active.

\item SUSY-GUT threshold effects.\\
  At the energy scale $\mugut$, threshold corrections induced by the
  non-degenerate SUSY-GUT spectrum have to be taken into account.
  Currently, only the one-loop corrections are
  available~\cite{Hall:1980kf,Weinberg:1980wa,Einhorn:1981sx}, which we include
  in our setup, although for consistency two-loop corrections would be necessary.

  A suitable linear combination of the three one-loop equations for $\alpha_i$
  in~(\ref{eq::dec}) and~(\ref{eq::gutdec}) leads to the following two
  relations
  \begin{eqnarray}
    4\pi\left(-\frac{1}{\alpha_1(\mu)}+3\,\frac{1}{\alpha_2(\mu)} - 2\,
      \frac{1}{\alpha_3(\mu)}\right)&=&-\frac{12}{5} L_{\mu H_c}
    \,,\nonumber\\
    4\pi\left(5\,\frac{1}{\alpha_1(\mu)}-3\,\frac{1}{\alpha_2(\mu)} - 2\,
      \frac{1}{\alpha_3(\mu)}\right)&=&-24\left( L_{\mu
      X}+\frac{1}{2}L_{\mu\Sigma}\right)\,,
    \label{eq::gutrel}
  \end{eqnarray}
  where $\alpha^{\rm SU(5)}$ has been eliminated.
  These equations allow for the prediction of the coloured triplet Higgs boson
  mass $\mhc$
  from the knowledge of the MSSM gauge
  couplings at the energy scale $\mu=\mugut$.
  It is furthermore common to define a new mass parameter
  $M_G=\sqrt[3]{M_X^2M_\Sigma}$, the so-called grand unified mass scale, that can   
  also be determined from the knowledge of the MSSM gauge
  couplings at $\mugut$.
  These observations makes it quite easy to test the
  minimal SUSY SU(5) model once the required
  experimental data are available in combination with a high-order analysis.

  For the Missing Doublet Model the above relations read
  \begin{eqnarray}
    4\pi\left(-\frac{1}{\alpha_1(\mu)}+3\,\frac{1}{\alpha_2(\mu)} - 2\,
    \frac{1}{\alpha_3(\mu)}\right)&=&-\frac{12}{5}\left( L_{\mu H_c}+L_{\mu
      H_{c'}}\right)+12\ln\frac{64}{3125} 
    \,,\nonumber\\
    4\pi\left(5\,\frac{1}{\alpha_1(\mu)}-3\,\frac{1}{\alpha_2(\mu)} - 2\,
    \frac{1}{\alpha_3(\mu)}\right)&=&-24\left( L_{\mu
      X}+\frac{1}{2}L_{\mu\Sigma}\right) - 12\ln\frac{262144}{1953125}
    \,,\nonumber\\
    \label{eq::gutrelmissdoub}
  \end{eqnarray}

\item Running from $\mugut$ to the Planck scale $M_{\rm Pl}$.\\ 
  The last sequence of our approach consists in
  the running within the SUSY-SU(5) model.
  We implemented the three-loop RGEs for the gauge~\cite{Jack:1996vg}, and the
  one-loop formulas for the Yukawa and Higgs
  self couplings~\cite{Hisano:1992jj} and impose that they can be described within 
  perturbation theory up to the Planck scale.

\end{enumerate}

\begin{table}[t]
  \begin{center}
  \renewcommand{\baselinestretch}{1.5}
    {\scalefont{0.8}
      \begin{tabular}{c|ccccc}
        & run             & dec                                & run 
        & dec                                           & run \\
        & $M_Z\to\mususy$ & $\alpha_i\to\alpha_i^{(\rm MSSM)}$ & $\mususy\to\mugut$
        & $\alpha_i^{(\rm MSSM)}\to\alpha^{(\rm GUT)}$  & $\mugut\to M_{\rm pl}$ \\
        \hline
        EW  & \hphantom{(3)}2(3) & \hphantom{(2)}1(2) & 3 &
        \raisebox{-.6em}{1(2)} &
        \raisebox{-.6em}{$3^\star$} \\[-.8em]
        QCD & 3 & 2 & 3 & \\
      \end{tabular}
    }
  \renewcommand{\baselinestretch}{1.0}
    \caption{\label{tab::run_dec}Loop corrections available for the individual
      steps of the running-decoupling procedure. The number is brackets
      indicate the loop-order needed for a consistent analysis.
      ($\star$ The running of the gauge and Yukawa coupling is performed to
      three- and one-loop accuracy, respectively.)
      }
  \end{center}
\end{table}

In Tab.~\ref{tab::run_dec} we summarize for the individual steps of the running-decoupling
procedure to which loop
order the perturbative corrections are currently available and implemented in
our setup.
The numbers in parenthesis indicate the loop order which would be necessary in
order to perform a fully consistent analysis with three-loop running and
two-loop decoupling. The phenomenologically most important ingredient
which is still lacking are the
two-loop decoupling relations at $\mugut$ as we will show in the next Section.

%- }}}

%- }}}
%- {{{ Phenomenological constraints on the SUSY GUT mass spectrum:

\section{\label{sec::pheno}Phenomenological constraints on the SUSY
  GUT mass spectrum}

\begin{figure}[t]
  \centering
  \begin{tabular}{cc}
    \includegraphics[width=.45\linewidth]{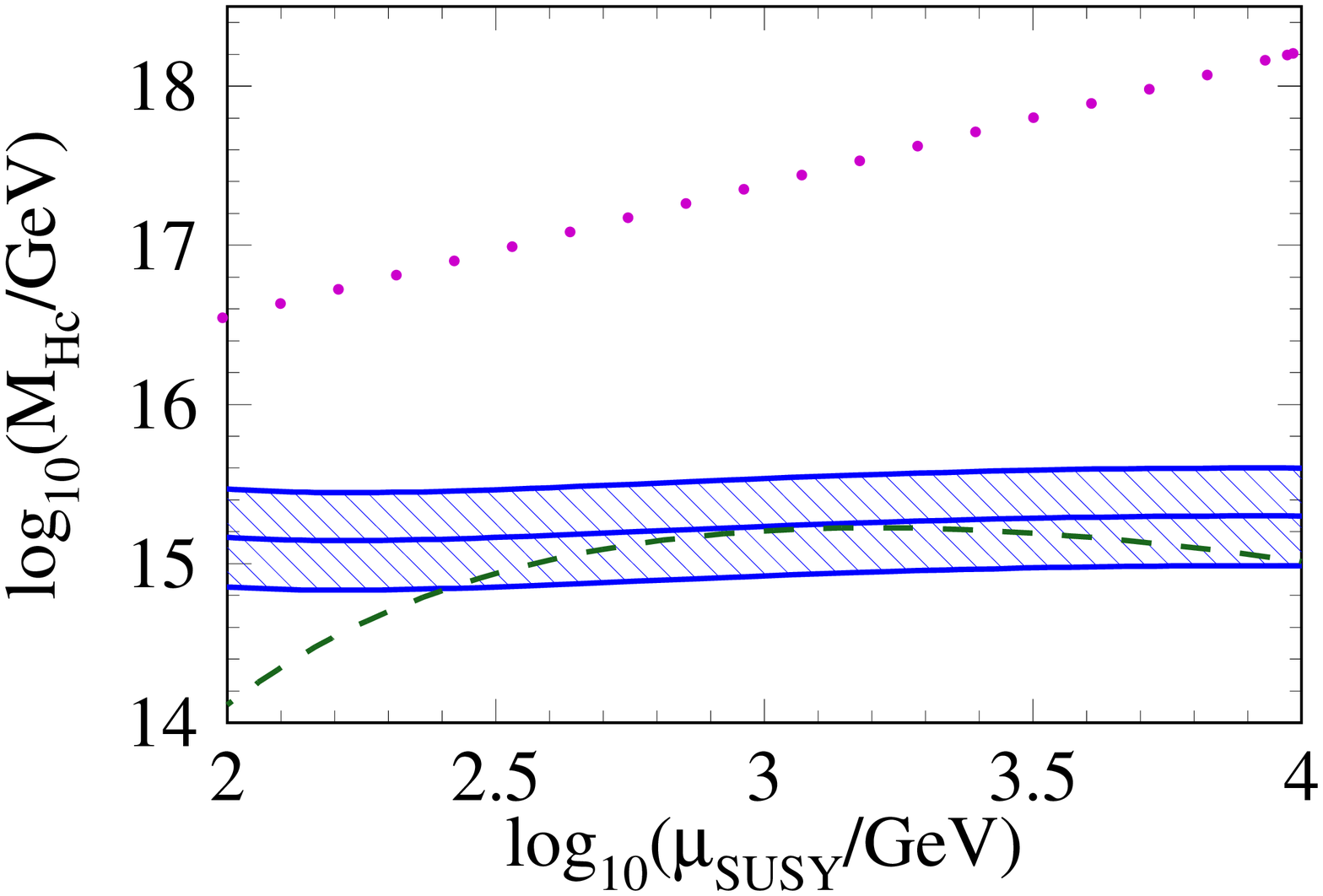}
    &
    \includegraphics[width=.45\linewidth]{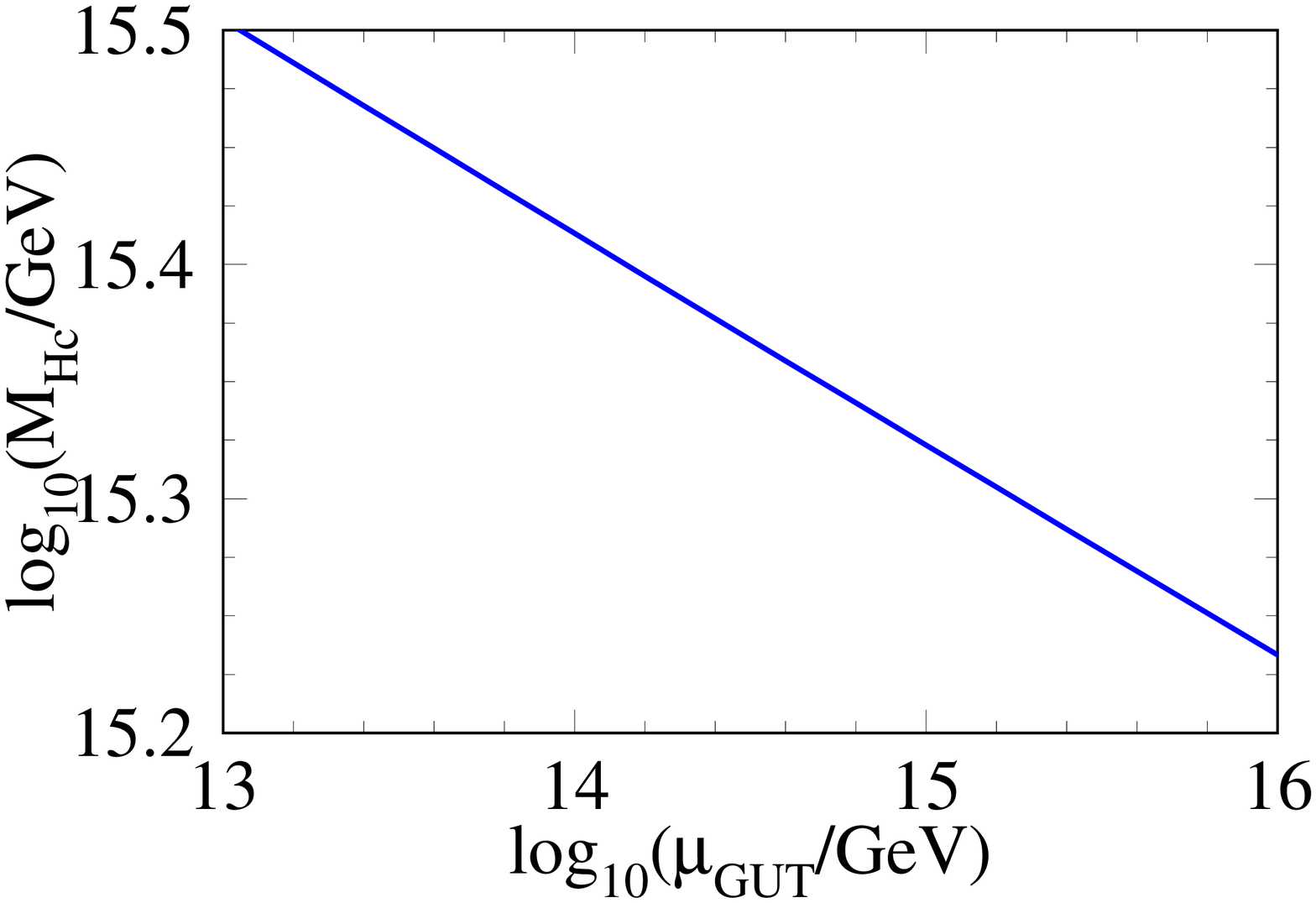}
    \\
    (a) & (b)
  \end{tabular}
  \caption{\label{fig::MHc_mudec}
    Dependence of $\mhc$ on $\mususy$ (a) and $\mugut$ (b)
    including successively higher orders. Dotted, dashed and solid lines
    correspond to the one-, two- and three-loop running analysis, respectively.}
\end{figure}

The running-decoupling prescription described in 
Section~\ref{sec::rundec} introduces two decoupling scales, $\mususy$
and $\mugut$, for 
the supersymmetric and GUT particles, respectively. These scales are not
determined by theory. However, on general grounds one expects that the 
GUT parameters become insensitive to the precise choice of these scales
when going to higher orders.
In Fig.~\ref{fig::MHc_mudec}(a) this is demonstrated for $\mhc$ considering
the dependence on $\mususy$ where the latter is varied between 100~GeV and
10~TeV. For illustration we adopt the 
{\tt mSUGRA} scenario  for the SUSY breaking mechanism with
\begin{eqnarray}
  m_0 &=&m_{1/2}\,\,=\,\,-A_0\,\,=\,\,1000~\mbox{GeV}\,,\nonumber\\
  \tan\beta &=& 3\,,\nonumber\\
  \mu&>&0\,,
  \label{eq::msugra_par1}
\end{eqnarray}
and generate with the help of {\tt SOFTSUSY}~\cite{Allanach:2001kg}
the supersymmetric mass spectrum. This results in squark masses which are of
the order to 2~TeV and thus in the upper range of what can be measured at the
CERN LHC. The dotted, dashed and solid lines in Fig.~\ref{fig::MHc_mudec}(a)
correspond to the one-, two- and three-loop running analysis, respectively, 
where the decoupling is performed at one order lower as required by consistency.
One finds a quite sizeable dependence of $\mhc$ at one-loop order varying
by almost two orders of magnitude. A significant reduction is observed
after inclusion of the two-loop effects leading to a variation of $\mhc$ by
only a factor two to three. Finally, after incorporating three-loop running
and two-loop matching corrections the variation
of $\mhc$ on $\mususy$ is about \mbox{$5\cdot 10^{14}$~GeV}
in the considered range for the
matching scale. It is furthermore remarkable that for $\mususy$ around
1000~GeV, which is close to the average of the supersymmetric masses, 
the two-loop corrections show a maximum and the three-loop corrections are
practically zero. In particular, they are significantly
smaller than for $\mususy=M_Z$ which has often been used as decoupling
scale for the supersymmetric particles. 

In Fig.~\ref{fig::MHc_mudec}(b) the dependence of $\mhc$ on $\mugut$ is
studied. A consistent three-loop analysis can not
be performed since the two-loop GUT matching relation is not yet
available. Nevertheless it is tempting to combine the three-loop running with
the one-loop matching effects which is represented by the solid line. One
observes that $\mhc$ varies by about $1.5\cdot10^{15}$~GeV.
This is of the same order of magnitude as the three-loop effect at the SUSY
scale if the matching is performed at $M_Z$ (see Fig.~\ref{fig::MHc_mudec}(a)).

% ----------

\begin{figure}[t]
  \centering
  \begin{tabular}{c}
    \includegraphics[width=.9\linewidth]{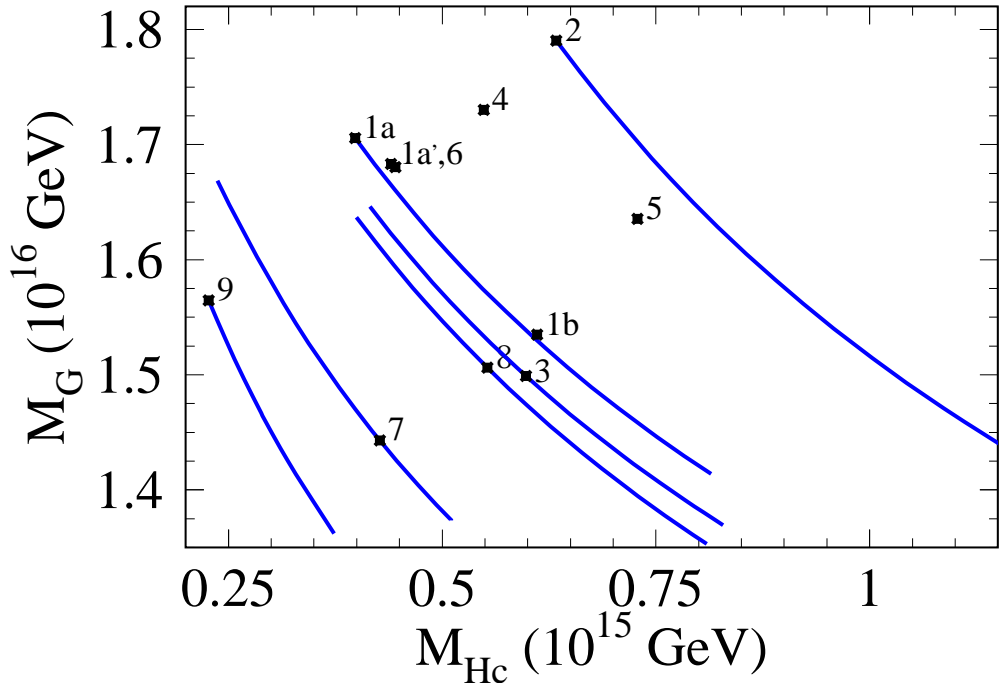}
  \end{tabular}
  \caption{\label{fig::mhc_mgut_sps}
    Dependence of $\mhc$ and $\mgut$ on the various SPS scenarios.}
\end{figure}

In the following we discuss the dependence of $\mhc$ and $\mgut$ on various
parameters entering our analysis. We start
with varying the supersymmetric mass spectrum by considering different
SPS scenarios~\cite{Allanach:2002nj,AguilarSaavedra:2005pw} and use
Eq.~(\ref{eq::gutrel}) in order to extract in each case both $\mhc$ and
$\mgut$. The decoupling scales are fixed to $\mususy = 1000$~GeV and $\mugut =
10^{16}$~GeV, respectively, which ensures, according to the previous discussion,
that the three-loop effect is rather small.
In Fig.~\ref{fig::mhc_mgut_sps} the results are shown in the $\mhc-\mgut$
plane where the lines indicate the slopes in case of SPS1a, SPS2, SPS3, SPS7,
SPS8 and SPS9.
Most scenarios lead to $\mhc$ masses between $0.25\cdot10^{14}$~GeV and
about $1\cdot10^{15}$~GeV: SPS9 (anomaly-mediated SUSY breaking)
gives the smallest and SPS2 the largest value of $\mhc$.

% ----------

\begin{figure}[t]
  \centering
  \begin{tabular}{cc}
    \includegraphics[width=.9\linewidth]{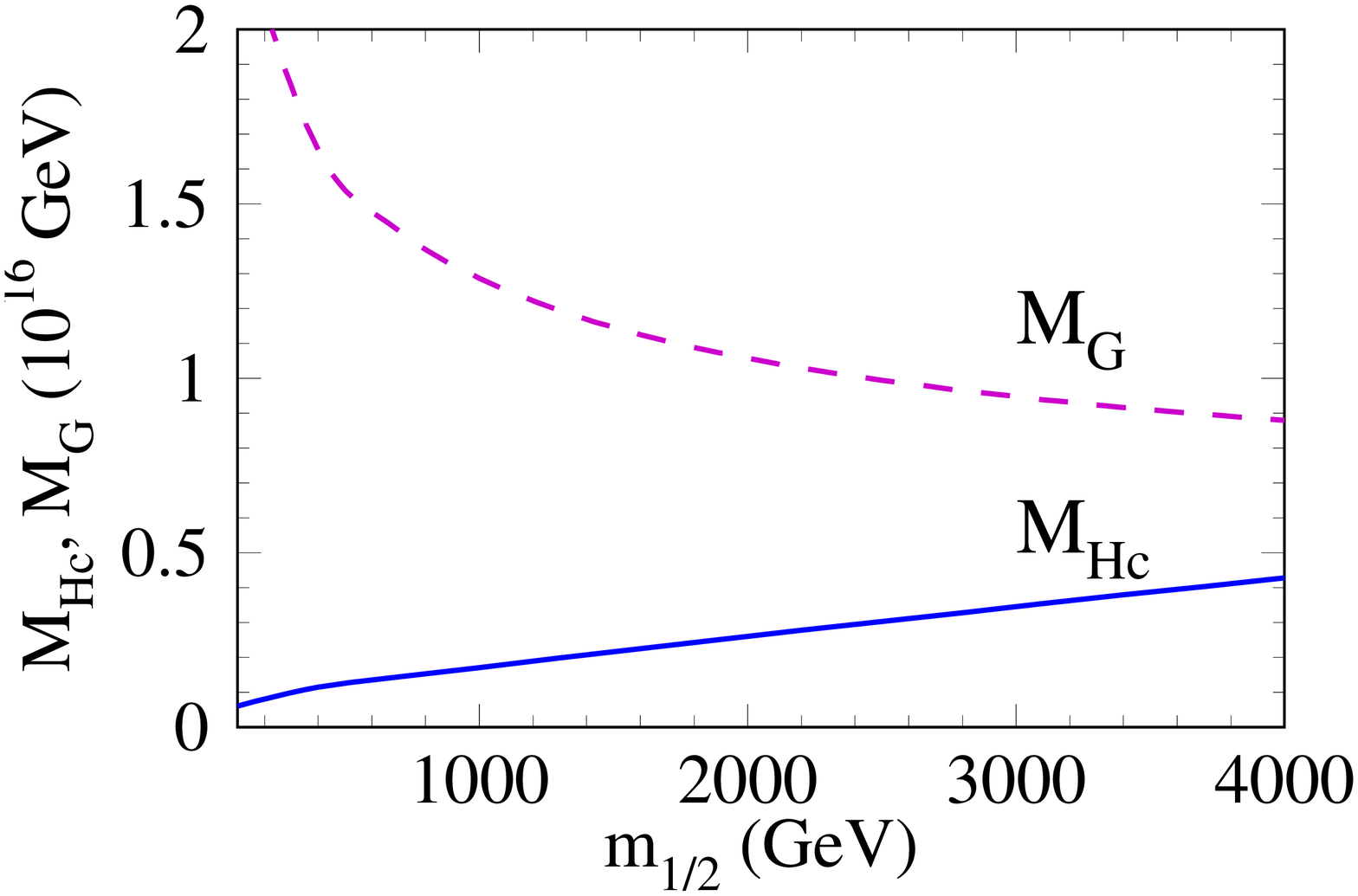}
  \end{tabular}
  \caption{\label{fig::mhc_msugra}
    Dependence of $\mhc$ (solid) and $\mgut$ (dashed) 
    on $m_{1/2}$.}
\end{figure}

A further illustration of the dependence of the GUT parameters on the SUSY
spectrum can be found in Fig.~\ref{fig::mhc_msugra} where we 
adopt the parameters of Eq.~(\ref{eq::msugra_par1})
and vary $m_{1/2}$. In this way we alter the SUSY spectrum
which enters the prediction of $\mhc$ and $\mgut$ via the
decoupling procedure at $\mu=\mususy$.
The solid and dashed lines correspond to $\mhc$ and $\mgut$, respectively,
which show a substantial variation. On the other hand, $m_0$, $\tan\beta$ and
$A_0$ have only a minor influence on the GUT masses and thus we refrain from
explicitly showing the dependence.

% ----------

In the following we study the effects of the uncertainties of the input
values $\alpha_i$ (c.f. Eq.~(\ref{eq::alpha123})) on $\mhc$ and $\mgut$. We fix
the SUSY spectrum as 
before (see Eq.~(\ref{eq::msugra_par1})) and set in a first step
$\mu_{\rm SUSY}=M_Z$ which has often been common practice in similar 
analyses (see, e.g., Ref.~\cite{Murayama:2001ur}).   
Taking into account correlated errors and performing a $\chi^2$
analysis leads to ellipses in the $\mhc-\mgut$ plane.
Let us mention that we can reproduce the results of
Ref.~\cite{Murayama:2001ur} after adopting their parameters and
restricting ourselves to the perturbative input used in that publication.

\begin{figure}[t]
  \centering
  \begin{tabular}{c}
    \includegraphics[width=.72\linewidth]{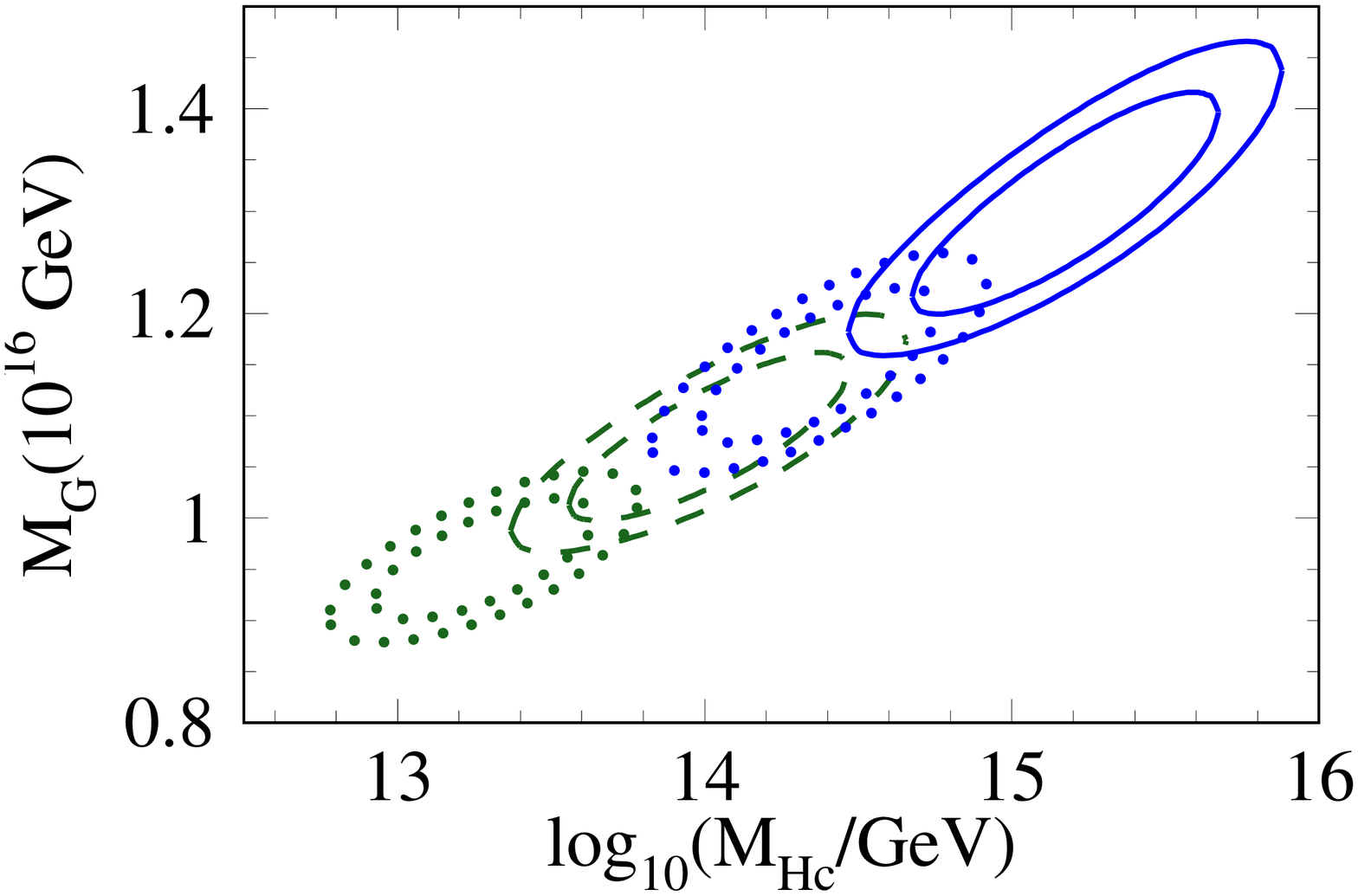}
    \\ (a) \\
    \includegraphics[width=.72\linewidth]{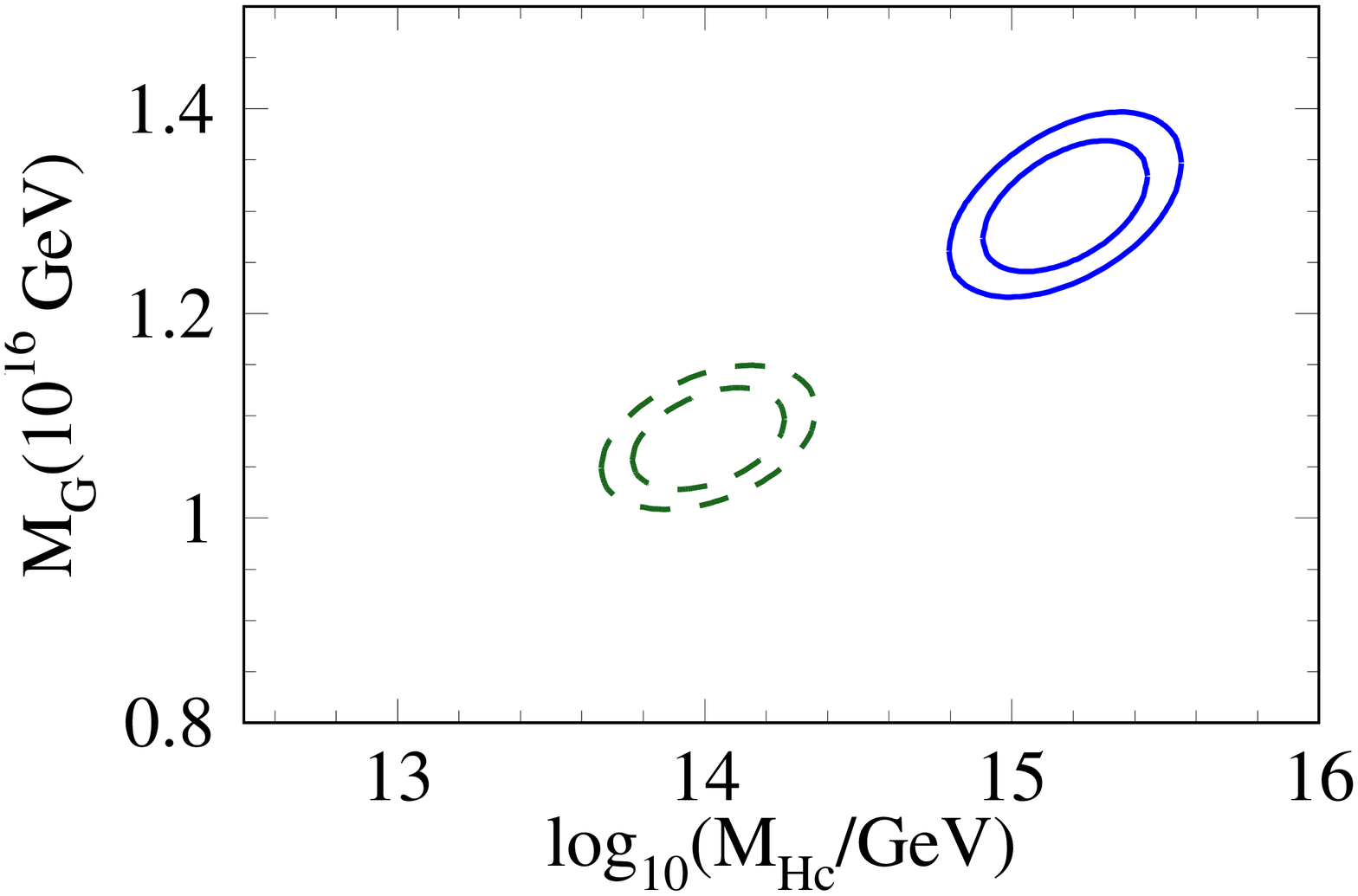}
    \\
    (b)
  \end{tabular}
  \caption{\label{fig::mhc_mgut_mz}
    Ellipses in the $\mhc-\mgut$ plane obtained from the uncertainties
    of the gauge couplings at the electroweak scale. In (a) the input parameters of
    Eq.~(\ref{eq::parin5}) have been used whereas in (b)
    $\delta\alpha_s=0.0010$ has been chosen.
    Dashed and solid lines correspond to the two- and three-loop
    analysis, respectively. The dotted lines in (a) have been obtained for
    $\alpha_s(M_Z)=0.1135\pm0.0014$ where the lower (upper) ellipse
    corresponds to the two- (three-) loop analysis.
  }
\end{figure}

In Fig.~\ref{fig::mhc_mgut_mz}(a) we show our results 
for the two- (dashed lines) and
three-loop (continuous lines)  analyses. In each case the two concentric
ellipses correspond to 68\% and 90\% confidence level, respectively, where
only parametric uncertainties from Eq.~(\ref{eq::alphasin}) have been taken
into account. 
A significant shift to higher masses of about 
an order of magnitude is observed for $\mhc$; $\mgut$ increases by
about $2\cdot 10^{15}$~GeV. This demonstrates the importance of the
two-loop matching and three-loop running corrections.
As has been  discussed in the context of Fig.~\ref{fig::MHc_mudec}
they are essential in order to remove the dependence on 
$\mu_{\rm SUSY}$. In fact, adopting in Fig.~\ref{fig::mhc_mgut_mz}
$\mu_{\rm SUSY}=1$~TeV would lead to ellipses for the two- and three-loop
analysis which were almost on top of each other and which would coincide with
the three-loop ellipses (solid lines) in Fig.~\ref{fig::mhc_mgut_mz}.

As expected, the uncertainty of $\alpha_s$ induces the
largest contributions to the uncertainties of $\mhc$ and $\mgut$. 
In particular, it essentially determines the semimajor axis of the ellipses.
Thus it is tempting to assign a more optimistic uncertainty of 
$\delta\alpha_s=0.0010$ and redo the previous analysis. The results,
shown in Fig.~\ref{fig::mhc_mgut_mz}(b), underline even more the
importance of the three-loop analysis. Whereas for
$\delta\alpha_s=0.0020$ the parametric uncertainty 
({\it i.e.} the size of the ellipses) and the shift due to
higher perturbative corrections are of the same order of magnitude,
in Fig.~\ref{fig::mhc_mgut_mz}(b) the latter is about twice
as big as the former.
Let us, however, stress once again that choosing $\mu_{\rm SUSY}$ close
to the supersymmetric mass scales leads to small three-loop effects
since the two-loop ellipses are essentially shifted on top of the
three-loop ones.

Recently there have been a few extractions of $\alpha_s$
based on higher order perturbative corrections with uncertainties
slightly above 1\%, which, however, obtain
central values for $\alpha_s$ close to 0.113
(see, e.g., Ref.~\cite{Gehrmann:2010rj}). 
Since these results are significantly lower than the value given in 
Eq.~(\ref{eq::parin5}) it is interesting to show 
in Fig.~\ref{fig::mhc_mgut_mz}(a) also the 
corresponding 68\% and 90\% confidence level ellipses (for the two- and
three-loop analysises) as dotted lines adopting
$\alpha_s(M_Z)=0.1135\pm0.0014$~\cite{Alekhin:2009ni}.  
One observes a big shift in the GUT masses,
in the case of $\mhc$ the central value is about one order of
magnitude lower than for the $\alpha_s$ value of
Eq.~(\ref{eq::parin5}). 

\begin{figure}[t]
  \centering
  \begin{tabular}{cc}
    \includegraphics[width=.45\linewidth]{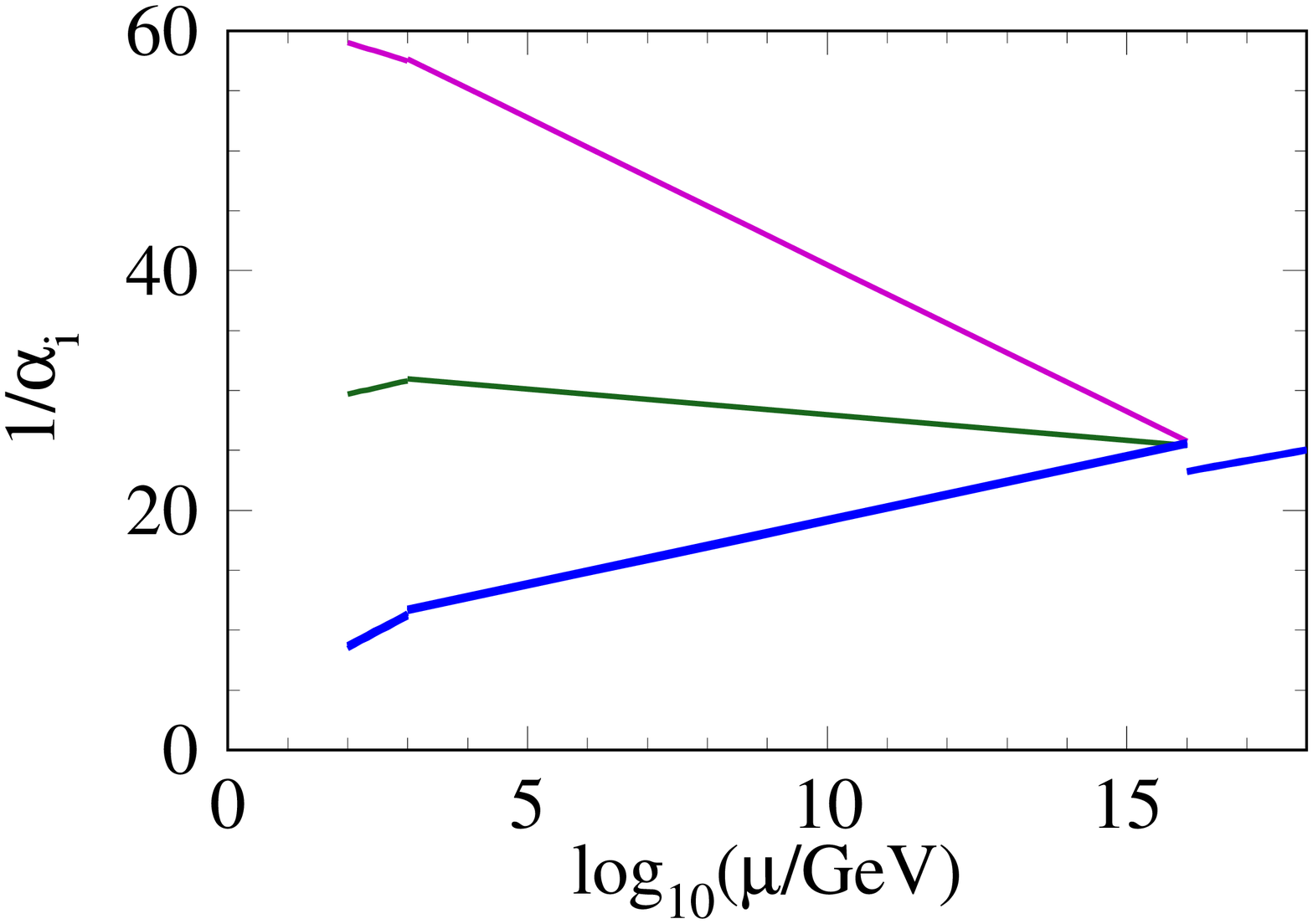}
    &
   \includegraphics[width=.45\linewidth]{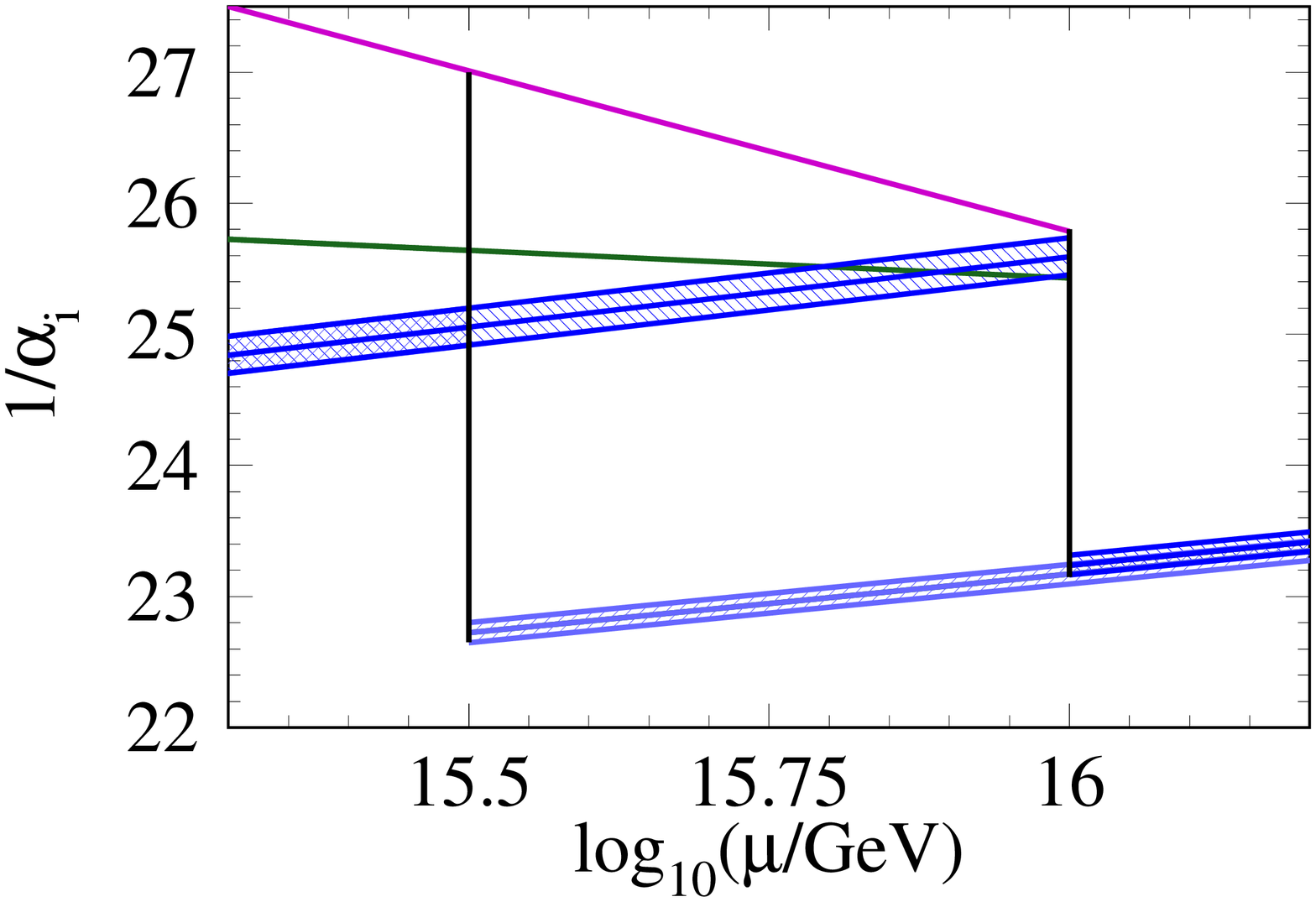}
    \\
    (a) & (b)
  \end{tabular}
  \caption{\label{fig::unification}Running of the gauge couplings from the
    electroweak to the Planck scale. The discontinuity for $\mu=\mususy$ and
    $\mu=\mugut$ are clearly visible. In panel (b) an enlargement of (a) for
    the region around $\mu=\mugut$ is shown where for the decoupling the two
    values $\mugut=10^{15.5}~\mbox{GeV}\approx 3.2\cdot 10^{15}$~GeV and
    $\mugut=10^{16}$~GeV have been chosen.} 
\end{figure}

In Fig.~\ref{fig::unification}
we visualize the running (and decoupling) of the gauge
couplings where the parameters of 
Eq.~(\ref{eq::msugra_par1}) together with 
$\mususy = 1000$~GeV and $\mugut = 10^{16}$~GeV have been adopted.
In addition we have chosen $M_\Sigma = 1\cdot 10^{15}~\mbox{GeV}$ which
leads to $\mhc = 1.7\cdot 10^{15}~\mbox{GeV}$ and $M_X = 4.6\cdot
10^{16}~\mbox{GeV}$. 
One can clearly see the discontinuities at the matching scales and the change
of the slopes when passing them.
In panel (b) the region around $\mu = 10^{16}$~GeV
is enlarged which allows for a closer look at the unification region.
The bands indicate $1\sigma$ uncertainties of $\alpha_i$ at the
electroweak scale (cf. Eq.~(\ref{eq::alphasin})).
In panel (b) we furthermore perform the decoupling of the super-heavy masses
for two different values of $\mugut$. One observes quite different threshold
corrections leading to a nice agreement of $\alpha^{\rm SU(5)}$ 
above $10^{16}$~GeV.
Fig.~\ref{fig::unification} stresses again that the uncertainty of $\alpha_s$
is the most important one for the   constraints that one can set on  GUT models from low-energy
data. Furthermore, it illustrates the size of the GUT threshold corrections
and emphasizes the importance of the two-loop corrections for the corresponding
decoupling constants.

\begin{figure}[t]
  \centering
  \begin{tabular}{c}
    \includegraphics[width=.75\linewidth]{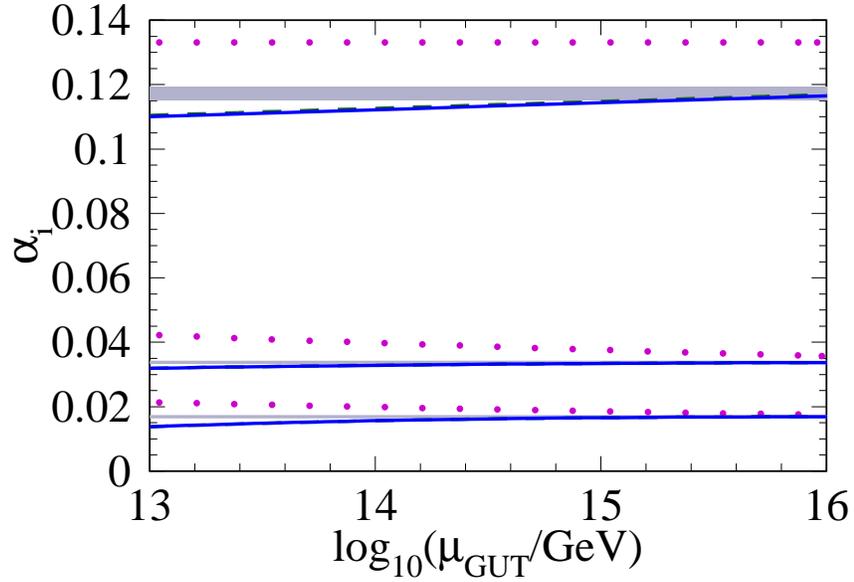}
    \\
    (a)
    \\
    \includegraphics[width=.75\linewidth]{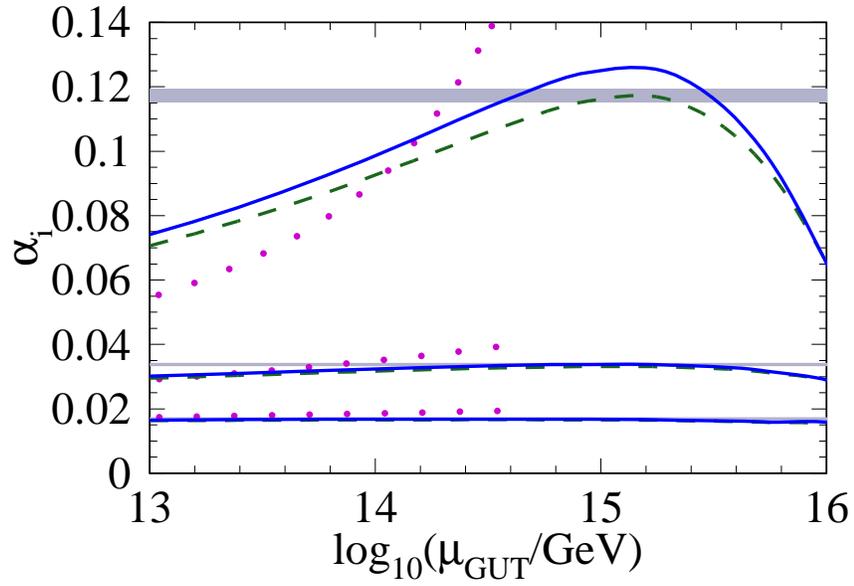}
    \\
    (b)
  \end{tabular}
  \caption{\label{fig::topdown}Gauge couplings at the weak scale as obtained by a
    top-down approach within minimal SUSY SU(5) (a) and the MDM (b) as a function
    of $\mugut$ (see text).
    In (b) the one-loop curves are only shown up to $\mugut\approx 4\cdot
    10^{14}$~GeV since beyond this scale $\alpha_3$ becomes quite large.}
\end{figure}

Finally we discuss the phenomenological consequences of our analysis in a
``top-down'' approach where we specify the parameters at the high scale and examine
the effect on the gauge couplings for $\mu=M_Z$.
For minimal SUSY SU(5) we choose the following parameters
\begin{eqnarray}
  \mhc &=& 3.67\cdot 10^{14}~\mbox{GeV}\,,\nonumber\\
  M_\Sigma &=& 2\cdot 10^{16}~\mbox{GeV}\,,\nonumber\\
  M_X &=& 1.58\cdot 10^{16}~\mbox{GeV}\,,\nonumber\\
  \alpha^{\rm SU(5)}(10^{17}~\mbox{GeV}) &=& 0.03986\,,
\end{eqnarray}
which guarantee the agreement of
$\alpha^{(6),\overline{\rm MS}}(M_Z)$,
$\sin^2\Theta^{(6),\overline{\rm MS}}(M_Z)$ and
$\alpha_s^{(6),\overline{\rm MS}}(M_Z)$ 
with their experimental counterparts for $\mugut=10^{16}$~GeV.
In Fig.~\ref{fig::topdown}(a) we fix $\mususy=500$~GeV and the SUSY spectrum
according to the SPS1a benchmark scenario which allows for a comparison with
the MDM (cf. Fig.~\ref{fig::topdown}(b)). We vary $\mugut$ by three orders of
magnitude where the dotted, dashed and solid line
correspond to the one-, two- and three-loop analysis of the described
procedure, respectively, and the bands reflect the uncertainties as given in
Eq.~(\ref{eq::alphasin}).
For this parameter choice we observe, as expected from the above
discussions, a sizeable two-loop effect but only a mild change after including
the three-loop corrections. Furthermore, the overall variation is small for
all three gauge couplings.
Note that the one-loop curve for $\alpha_s$ is independent of $\mugut$ since
the one-loop coefficients of the $\beta$ functions in SUSY QCD and minimal
SUSY SU(5) are identical.

In Fig.~\ref{fig::topdown}(b) we perform the same analysis within the MDM
adopting SPS1a,\footnote{Adopting the parameters from Eq.~(\ref{eq::msugra_par1})
  leads to a non-perturbative values of the gauge coupling at the Planck
  scale.}
$\mususy=500$~GeV and the parameters\footnote{The relatively
  large value of $\alpha^{\rm SU(5)}(10^{17}~\mbox{GeV})$ is required due to
  the large Casimir constants entering the $\beta$ function above the GUT
  scale.}
\begin{eqnarray}
  \mhc &=& 6\cdot 10^{18}~\mbox{GeV}\,,\nonumber\\
  \mhcp &=& 1\cdot 10^{16}~\mbox{GeV}\,,\nonumber\\
  M_\Sigma &=& 2\cdot 10^{15}~\mbox{GeV}\,,\nonumber\\
  M_X &=& 3\cdot 10^{16}~\mbox{GeV}\,,\nonumber\\
  \alpha^{\rm SU(5)}(10^{17}~\mbox{GeV}) &=& 0.1504\,.
\end{eqnarray}
One observes a much stronger variation of the predicted values of $\alpha_i$,
in particular in the case of the strong coupling which varies between 0.06 and
0.13 in the considered range of $\mugut$.
This demonstrates the importance of the two-loop threshold corrections which
are expected to significantly reduce the dependence on $\mugut$.

%----------

At this point a discussion about the additional constraint on the Higgs
triplet mass 
$M_{H_c}$ that can be derived from the non-observation of the proton
decay is in order. The latest upper bound on the proton 
decay rate for the channel $p\to K^+\bar{\nu}$~\cite{Kobayashi:2005pe}
is $\Gamma_{\rm exp}=4.35\times 10^{-34}/\mbox{y}$. In order to
translate it into a lower bound for the  the Higgs triplet mass, one
needs an additional assumption about the Yukawa couplings that enter
the expression of the decay rate $\Gamma(p\to K^{+}\bar{\nu})$.  As
pointed out in Ref.~\cite{EmmanuelCosta:2003pu} this is because down quark
and lepton Yukawa couplings fail to unify within the minimal renormalizable SUSY
SU(5) model and so a completely consistent treatment is not possible.
Therefore one could either choose\footnote{$Y_{ql}$ is the Yukawa
  coupling of the quark and lepton doublet to the Higgs colour triplet.
  $Y_{ud}$ is the corresponding coupling for the up and down quark singlet.} (i) $Y_{ql}=Y_{ud}=Y_d$ or
(ii) $Y_{ql}=Y_{ud}=Y_e$, which leads to completely different
phenomenological consequences. 
We are aware that both cases are equally justified
once higher dimensional operators are included. Since these operators
further weaken the bounds presented below, we refrain to include them
into the analysis in this paragraph.
For the case (i) and sparticle masses around 1~TeV the lower bound
for  the Higgs triplet mass can be read off from Fig.~2 of
Ref.~\cite{EmmanuelCosta:2003pu} 
and amounts to  $M_{H_c}\ge 1.05 \times 10^{17}$GeV whereas for
the second choice it becomes $M_{H_c}\ge 5.25\times 10^{15}$GeV.
From our phenomenological analysis 
presented above it turns out that within the minimal
SUSY SU(5) model the upper bound for $M_{H_c}$ is 
of about $10^{16}$GeV.  So, the substantial increase   of about one
order of magnitude for the upper bound on $M_{H_c}$
induced by the three-loop order running analysis attenuates
the tension
between the theoretical predictions made under the assumption (i) and the
experimental data. The choice (ii) for the Yukawa couplings  clearly shows that
the minimal SUSY SU(5) model cannot be ruled out by  the current
experimental data on proton decay rates.   More experimental
information about the SUSY mass spectrum and proton decay rates is
required in order to be able to draw a firm conclusion.

On the other hand, it is commonly recognized that SUSY SU(5) in its
minimal version cannot provide  
the underlying model for a consistent GUT because of the failure of Yukawa unification. 
As mentioned in the Introduction, one of the most appealing solutions of this
problem is the inclusion  of  $M_{\rm Pl}$ scale effects  
using the effective theory approach, {\it i.e.}, including higher 
dimensional operators. Explicitly, this translates into~\cite{Bajc:2002pg}
\begin{eqnarray}
M_{(8,1)} \ne M_{(1,3)} \quad \mbox{and} \quad M_{H_c}=M_{H_c}^{(0)}
 \left(\frac{M_{(1,3)}}{M_{(8,1)}}\right)^{\frac{5}{2}}\,,
\end{eqnarray} 
where $M_{H_c}^{(0)}$ denotes the mass of the Higgs triplet
in the minimal renormalizable model. In this case, a suitable choice of the
Yukawa matrices can  significantly reduce the proton decay amplitudes, so that
values for $M_{H_c}$ of the order $\mathcal O (10^{15}\mbox{GeV})$ even for
larger values of $\tan\beta\simeq 15$ are allowed. So the prediction of the
non-renormalizable version of the SUSY SU(5) model is well under the
experimental upper bound. However, the inclusion of new undetermined
parameters together with the higher dimensional operators makes the tests
of such models much more complicated.

At this point we try to compare with the findings of the
analysis~\cite{Dorsner:2006ye} based on the non-renormalizable version
of minimal SU(5) where the bound $\mhc > 3.7\times10^{17}$~GeV is
given. This is the most stringent bound\footnote{See 
Ref.~\cite{Dorsner:2006ye} for the exact definition of this bound.}
that can be derived from current proton decay experimental data.
In order to be consistent with our upper bound on $\mhc^{(0)}$
(cf. Fig.~\ref{fig::mhc_mgut_mz}) of about $10^{16}$~GeV one requires
$M_{(1,3)}/M_{(8,1)} \gtrsim  4.2$ which is a quite moderate value.

%- }}}
%- {{{ Conclusions:

\section{\label{sec::concl}Conclusions}

Already in the early days of the Standard Model there have been studies of
theories which predict the unification of the coupling constants at high
energies. Since it is 
not possible to reach such energies in collider experiments it is
necessary to establish relations between the couplings at the
electroweak scale, where precise measurements are available, and the
corresponding quantities at the unification scale. Next to the beta functions
covering the running also threshold corrections at the SUSY
and GUT scale constitute crucial input.

We have considered the so-called minimal  supersymmetric SU(5) GUT theory and
have studied the gauge coupling unification applying one-, two- and three-loop
running. The main effect of the three-loop running (accompanied by two-loop
decoupling 
relations) is the stabilization w.r.t. the variation of the decoupling scales.
In general sizable three-loop effects are observed if the decoupling scale for
the supersymmetric particles is not chosen in the vicinity of the masses of the
supersymmetric particles.
In particular, for $\mu_{\rm SUSY}=M_Z$, which is the canonical choice often
adopted in the literature, one observes an increase of the coloured Higgs
triplet mass by about an order of magnitude. 
Already in previous studies it has been shown that the
non-renormalizable version of minimal SUSY SU(5) cannot be excluded
by the experimental bound on the proton decay
rate~\cite{Bajc:2002bv,Bajc:2002pg,EmmanuelCosta:2003pu}.
Our results attenuate the tension between the theoretical predictions
of $\mhc$ and the experimental results even more.

Our analysis includes all the state-of-the-art theoretical input. It could be
improved by including the two-loop GUT
threshold corrections which are not yet available.
This induces an uncertainty
of 
about 0.3 on $\log_{10}(\mhc/\mbox{GeV})$ which constitutes the major
theory-uncertainty. 
We  observe that the effects of the  two-loop GUT threshold corrections
become particularly important in the Missing Doublet Model where also larger
effects can be observed.
As far as the parametric uncertainty is concerned we mention the dependence
on the supersymmetric mass spectrum and the uncertainty on $\alpha_s$. In both
cases a shift of $\mhc$ of a few times $10^{15}$~GeV is observed.

The most popular extensions of the minimal SUSY SU(5) model, the Missing Doublet Model
and the non-renormalizable version of the
SUSY SU(5) model  comprising Planck-scale operators, can not be excluded using 
only the currently available theoretical
and experimental data. They either contain additional free parameters as compared to
the minimal model or are affected by large theoretical uncertainties, so that no firm
conclusion can be drawn for them.

%- }}}

%- {{{ Acknowledgements:

\section*{Acknowledgements}
We would like to thank Miko{\l}aj Misiak and Robert Harlander
for carefully reading the manuscript
and many useful comments. We are grateful to 
Thomas Teubner and Daisuke Nomura for communications regarding
$\Delta\alpha_{\rm had}^{(5)}$.
This work was supported by the DFG through SFB/TR~9 ``Computational
Particle Physics'', the Graduiertenkolleg ``Hochenergiephysik und
Teilchenastrophysik'', 
the ``Studienstiftung des Deutschen Volkes'' and the
``Landesgraduiertenf\"orderung'' of the state of Baden-W\"urttemberg.

%- }}}
%- {{{ Bibliography:

%- }}}

\end{document}